\documentclass[12pt]{article}
\usepackage[utf8]{inputenc}
\usepackage[T1]{fontenc}
\usepackage[english]{babel}
\usepackage{amsmath}
\usepackage{amssymb}
\usepackage{graphicx}
\usepackage{float}
\usepackage{multirow}
\usepackage{natbib}
\usepackage{pgfplots}
\usepackage{multirow}
\usepackage[title,titletoc,toc]{appendix}
\usepackage{eurosym}
\usepackage{booktabs}
\usepackage{color,soul}
\usepackage{pbox}
\usepackage{comment}
\usepackage{floatpag}
\usepackage{subcaption}

\usepackage{hyperref}

\newcommand{\nascondi}[1]{}

\title{COVID-19: $R_0$ is lower\\
where outbreak is larger}

\date{This version: \today}

\author{Pietro Battiston\thanks{Department of Economics and Management,
University of Parma.
ORCID: 0000-0001-5343-7861.
me@pietrobattiston.it}, Simona
Gamba\thanks{Department of Economics and Finance, Catholic University of Milan.
ORCID: 0000-0002-5538-2873.
gamba.simona@gmail.com}}

\begin{document}

\maketitle

\vspace{-1cm}

\begin{abstract}
We use daily data from Lombardy, the Italian region most affected by the
COVID-19 outbreak, to calibrate a SIR model individually on each municipality.
These are all covered by the same health
system and, in the post-lockdown phase we focus on, all
subject to the same social distancing regulations.
We find that municipalities with a higher number of cases at the beginning of
the period analyzed have a lower rate of diffusion, which cannot
be imputed to herd immunity. In particular,
there is a robust and strongly significant negative correlation between the
estimated basic reproduction number ($R_0$) and the initial outbreak size,
in contrast with the role of $R_0$ as a \emph{predictor} of outbreak
size.
We explore different possible explanations for this phenomenon and conclude that
a higher number of cases causes changes of
behavior,
such as a more strict adoption of social distancing measures among the
population, that reduce the spread.
This result calls for a transparent, real-time distribution of
detailed epidemiological data, as such data affects the behavior of populations
in
areas affected by the outbreak.
\end{abstract}
\newpage

\section{Introduction}\label{intro}

The basic reproduction number, or $R_0$, represents the average
number of secondary cases produced by a single
infected case in an otherwise susceptible
population, and it is typically used as a reference value
to assess the transmissibility of an infectious disease in a given population.
Given a number of individuals susceptible to infection, a disease with higher
$R_0$ will infect a larger number of individuals.
There is hence an obvious positive relationship between the $R_0$ and the
resulting size of an outbreak \citep{tildesley2009r0}.

However, the value of $R_0$ during an outbreak does not only depend on ex-ante
features of a
virus or a population, but potentially also on the \emph{response} of
both population and authorities to the outbreak.
This is particularly true in the context of the COVID-19 pandemic, to which most
countries in the world have reacted with some form of social distancing
measures, or lockdown. In absence of a vaccine or effective drugs,
these measures are the best weapon to reduce the number of deaths, as well as 
the number of intensive care unit beds required
\citep{kucharski2020early,flaxman2020estimating,Ferguson,Greenstone}.

In the present study, we analyze data on the diffusion of COVID-19 in Lombardy,
the region of Italy most heavily affected by the pandemic (\citealp{d2020early}
provide an accurate description of the early phase of the outbreak in such
region).
Specifically, we employ daily data on the number of individuals positive to
COVID-19 at the municipality level, focusing on a period in which the entire
country was subject to a lockdown.
All municipalities under analysis share the same public health system and, in
the period considered, were subject to the same social distancing regulation.
However, at the start of the period, they were characterized by a strong
heterogeneity in the number of cases, both in absolute terms and in terms of
cases per capita.

We study a period beginning on March 25, 2020, that is, more
than two weeks after the lockdown regulation was put in place, and ending with
April 14, 
when such regulations still held: this means that
movements across municipalities are severely
restricted, requiring any travelers to present a valid (typically work or
health related) justification for their journey.

We fit a Susceptible-Infected-Recovered (SIR) model on data from each
municipality and
find that the estimated $R_0$ is negatively correlated with the prevalence in
the municipality at the beginning of our period.
This result holds both when considering the absolute and per capita number of
cases and is robust to different specifications and sample disaggregations.

We present and compare different complementary explanations for this finding.
Early and widespread testing 
increases the reported number of cases and might allow the 
authorities to slow the spread of the pandemic by isolating known cases.
At the same time, where the number of cases is 
higher, the population might comply more strictly with the lockdown measures, thus
reducing 
the rate of spread: we show in Section \ref{sec:interp} why the latter mechanism
is most likely to drive our results.

\section{Data}

We employ count data of per-municipality recorded cases, updated daily and
distributed by regional authorities.
We do not rely on data on recovered and deceased individuals, as such data are
not available with the required geographical disaggregation.

\begin{figure}
    \centering
    \caption{Distribution of cases and cases increase}\label{fig:lomb-data}
    \begin{subfigure}[b]{10cm}
    \begin{center}
        \includegraphics[width=10cm]{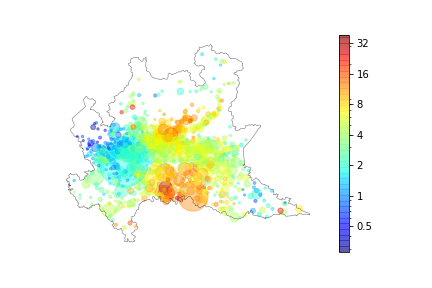}
    \caption{Cases on March 25}
    \label{fig:data-25}
    \end{center}
    \end{subfigure}
    ~
    \begin{subfigure}[b]{10cm}
    \begin{center}
        \includegraphics[width=10cm]{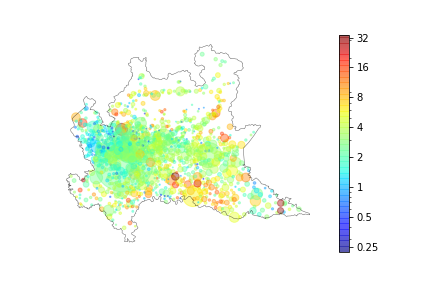}
    \caption{New cases between March 25 and April 14}
    \label{fig:data-diff}
    \end{center}
    \end{subfigure}

\raggedright \emph{Note:} dot size represent absolute numbers, colors 
represent cases per one thousand inhabitants.
\end{figure}

Data are available starting from March 25, 2020 and cover a period of twenty-one 
days during which lockdown measures were always in place.
We verify that only minimal deviations appear
between regional data  and the aggregation of municipal data.
Out of 1507 municipalities in Lombardy,  960 had at least
one recorded COVID-19 case as of this date.
Figure \ref{fig:data-25} displays the number of cases (size of the dots) and 
the cases 
per capita (color of the dots) as of March 25 for each of these 960 municipalities.
Similarly, Figure \ref{fig:data-diff} displays the number of new cases (size of 
the dots) and the number of new cases per capita (color of the dots)
recorded in each municipality between March 25 and April 14. 

It should be noted that official data concerning the COVID-19 outbreak in Italy
has been found to be strongly incomplete, both in terms of positive
individuals and of casualties: several researchers have estimated an outbreak
size much higher than that suggested by official numbers
\citep{flaxman2020estimating}, while others have corroborated this with an
analysis of anomalies in death
rates.\footnote{https://www.lavoce.info/archives/65042/decessi-da-covid-facciamo-chiarezza-sui-dati-istat/}
Moreover, local testing strategies are known to have deviated from WHO guidelines
and have
changed over time, also depending on available resources:
towards the end of
our period of interest, more subjects with mild symptoms were tested.
For this reason, some 
researchers have put forwards adaptations of the SIR model that account for
a threshold in the capacity of the health system.\footnote{https://www.lavoce.info/archives/65036/perche-e-cosi-alta-la-mortalita-da-coronavirus-in-lombardia}
Such problems are not specific to Italy, as official data from a number of
countries have been questioned.
More in general, the difficulty in obtaining reliable data on the number of
infected, deceased and recovered individuals calls for refinements of
traditional epidemiologic models \citep{Atkeson,FBK}.

Given our research question, these caveats are of limited importance.
Indeed, the focus of the present work is to document \emph{differences} in response
across municipalities, rather than to precisely estimate the epidemiological
parameters or expected duration of the COVID-19 outbreak in Lombardy.

Data on population size is obtained from the Italian National Istitute of
Statistics (ISTAT).

\section{Methods}
\label{sec:methods}

We fit a SIR model on each municipality in the period of twenty-one days
beginning in March 25.
Given the short time span considered, we employ a simplified SIR model which
does not account for natural rate of mortality.
Hence, the model is entirely defined by setting few parameters: $\beta$, which
determines the rate at which susceptible ($S$) individuals become infected ($I$);
$\gamma$, which determine the rate at which infected individuals become
recovered ($R$); the 
initial number of infected and
recovered individuals, and the population size $N$ ($=S+I+R$).
We take population size from official statistics.
We hence consider a discretized version of the continuous SIR model -- each
period corresponding to a day -- and automatically explore the parameter space
for $\beta$, $\gamma$, and
the initial value for $I$ and $R$, looking for the combination that provides the best
fit.\footnote{While in principle we could consider a constraint by which the sum
of the initial values of $I$ and $R$ adds up to the initial number of cases,
this is not required nor optimal. It is not required because the fitting
procedure minimizes the fit error in \emph{all} periods, including the first;
it is not optimal because the first datum might legitimally be affected by
fluctuations that deserve no larger importance than subsequent ones.}
Specifically, the goodness of fit is maximized by minimizing the sum of square
residuals between the cases count and the sum of the $I$ and $R$ pools
sizes.\footnote{The optimization algorithm is described in the Supplementary
Information.}
The initial values for the free parameters are set to those calibrated on the
entire Lombardy region.

Given that the SIR model assumes a
non-null initial population of infected individuals, we only consider the 960
municipalities satisfying this condition.
We further drop 47 municipalities which had new cases recorded on only one or
two dates, hence reducing to 
913 
municipalities.\footnote{The fitting procedure may become unreliable if provided
too few updating points; in particular, a linear growth of cases yields an
indeterminacy problem whereas a similar prediction can be obtained with very
different paramter values.}
Although this sample selection might in
principle affect our results, we show in Section \ref{sec:sensitivity} that this 
is not the case.

\begin{figure}
    \centering
    \caption{Comparison of fitted SIR model and total cases count}\label{fig:data-fits}
    \begin{subfigure}[b]{4.3cm}
    \begin{center}
        \includegraphics[width=4.4cm]{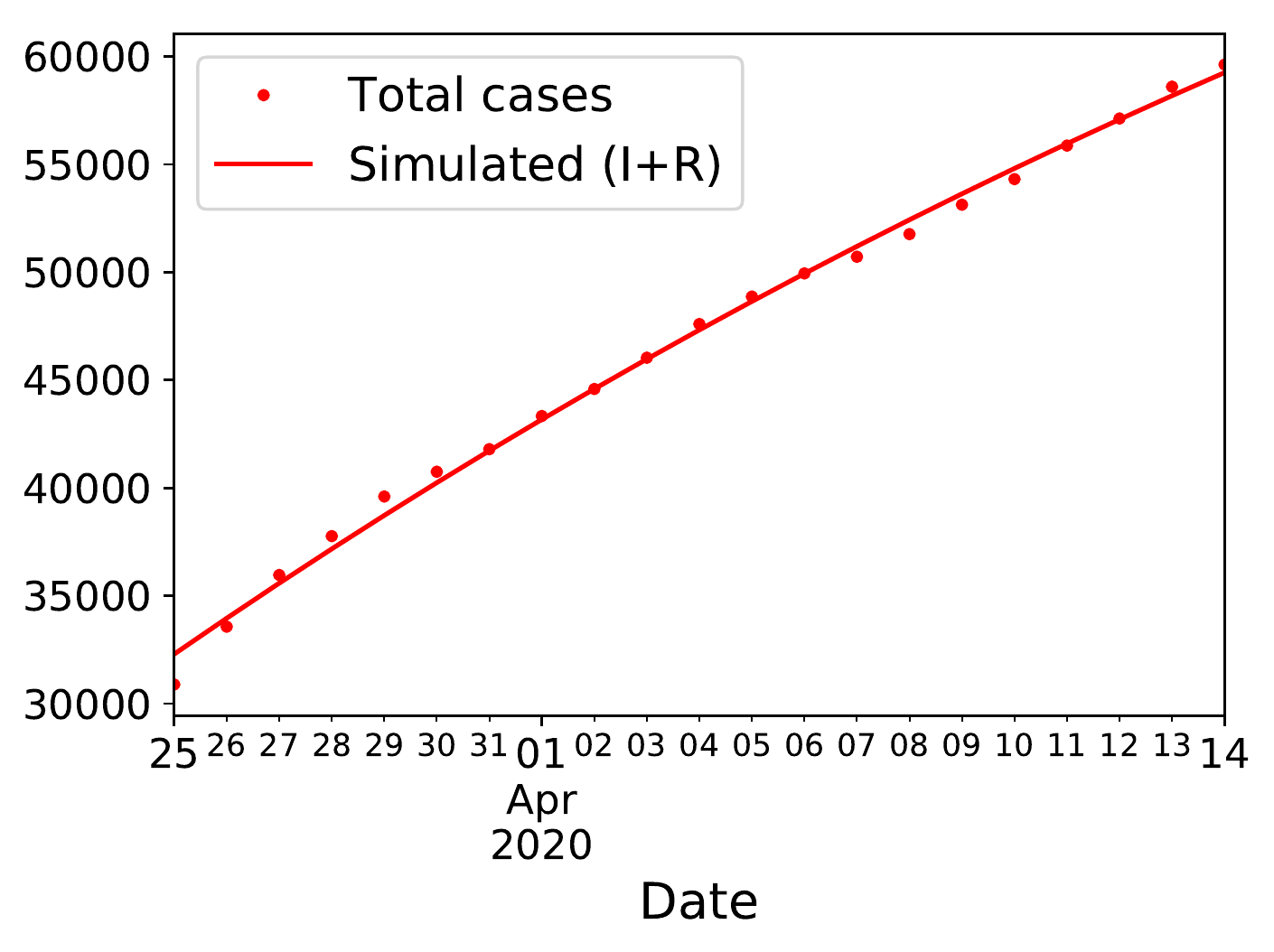}
    \caption{Lombardy}
    \label{fig:data-fit-lombardy}
    \end{center}
    \end{subfigure}
    ~ 
    \begin{subfigure}[b]{4.3cm}
    \begin{center}
        \includegraphics[width=4.4cm]{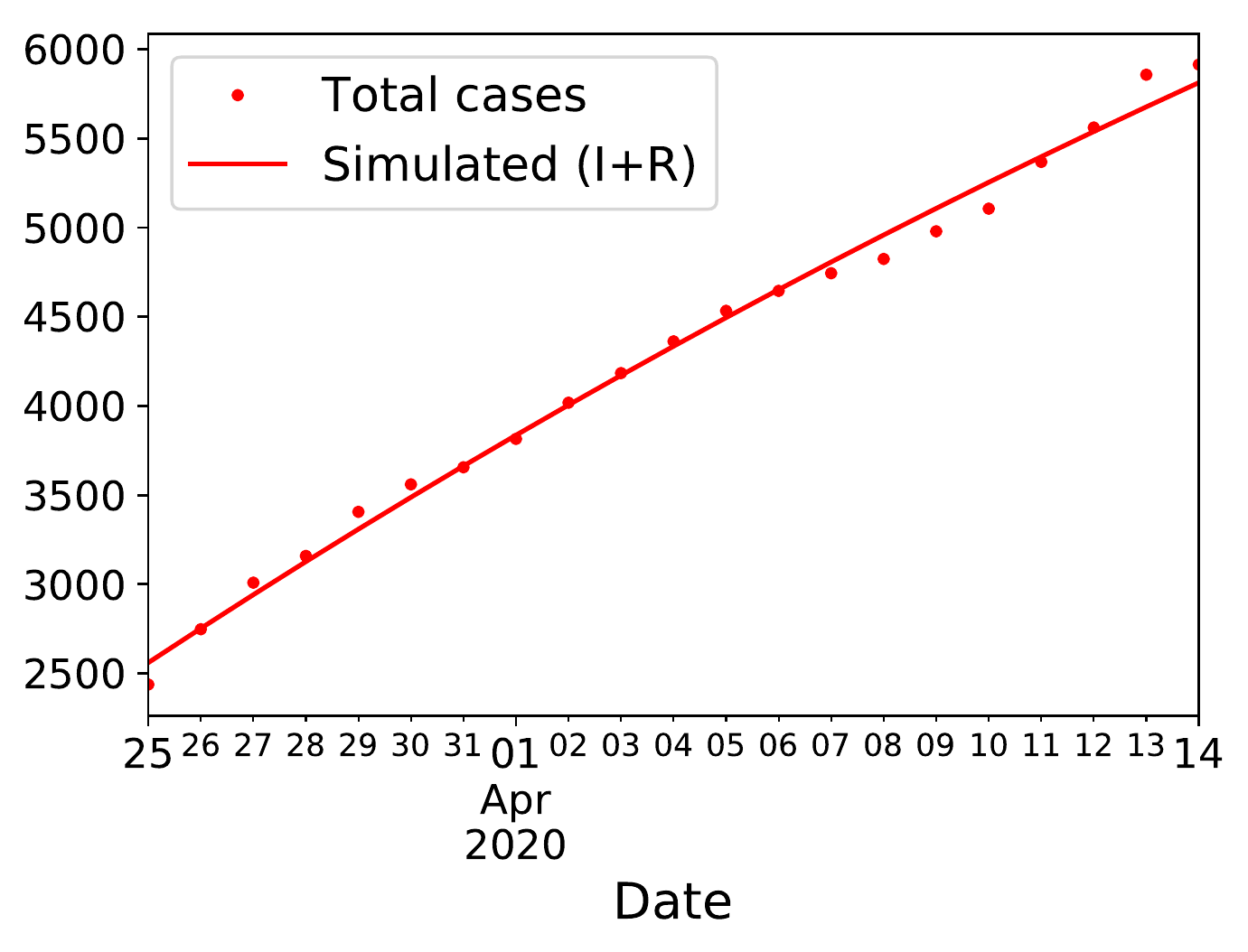}
    \caption{Milan}
    \label{fig:data-fit-milano}
    \end{center}
    \end{subfigure}
    ~ 
    \begin{subfigure}[b]{4.3cm}
    \begin{center}
        \includegraphics[width=4.4cm]{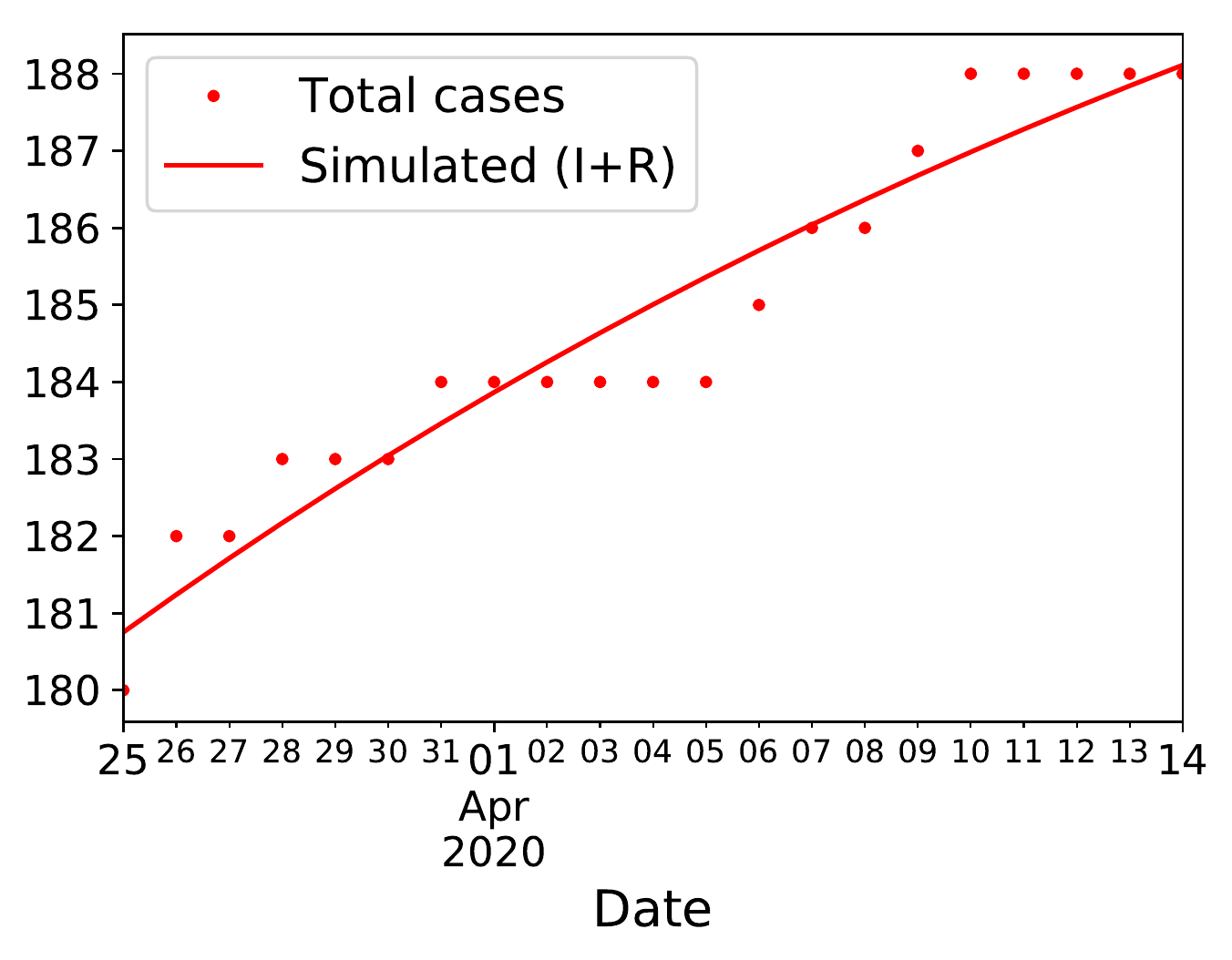}
    \caption{Castiglione d'Adda}
    \label{fig:data-fit-castiglione}
    \end{center}
    \end{subfigure}
\raggedright {\footnotesize \emph{Note:} fit between data and the corresponding 
SIR model for Lombardy region (left) and the most affected
municipalities at the beginning of our period of interest in absolute and per 
capita terms, respectively (center, right).}
\end{figure}

Figure \ref{fig:data-fit-lombardy} displays the fit between data 
at the regional level and the corresponding simulated SIR model. Figures  
\ref{fig:data-fit-milano} and \ref{fig:data-fit-castiglione} are the equivalent 
for Milan and Castiglione d'Adda:
these are the two municipalities which, at the 
beginning of our period of
interest, had been most heavily hit in 
absolute and per capita terms, respectively.
Note that a weekly fluctuation can be observed for all municipalities: this is
in line with documented evidence that less tests are processed during the
weeekend, and the effect reverberates on the number of positive detected cases 
with a delay of two to three days. We expect these fluctuations to affect the
entire
region homogeneously.

Once we find the best SIR parameters for each municipality, we regress the
estimated $R_0$ (the ratio of the estimated $\beta$ and $\gamma$) on the outbreak size
within the
municipality as of March 25.
We focus on the \emph{per capita} number of cases, as we expect any effect to
be related to the \emph{prevalence} of the outbreak -- a same
number of cases will be perceived in a very different way in Milan or in a small
municipality.

\section{Results}
\label{sec:res}

\begin{figure}
    \begin{center}
    \caption{Distribution of $R_0$}\label{fig:R0-dist}
    \label{fig:r0-dist}
    \begin{subfigure}[b]{7cm}
        \includegraphics[height=5.2cm]{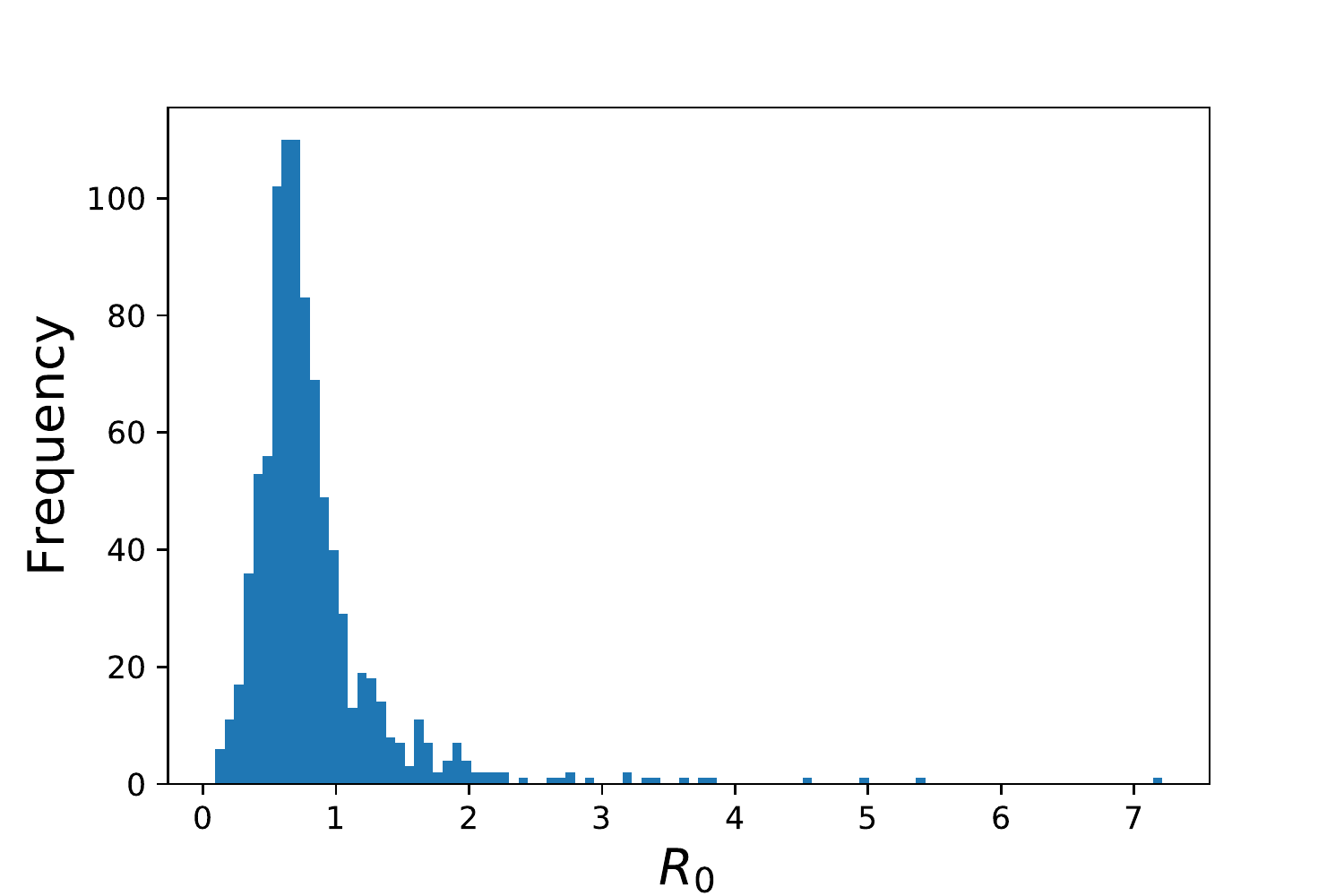}
    \end{subfigure}
    \begin{subfigure}[b]{6.5cm}
        \includegraphics[height=5.2cm]{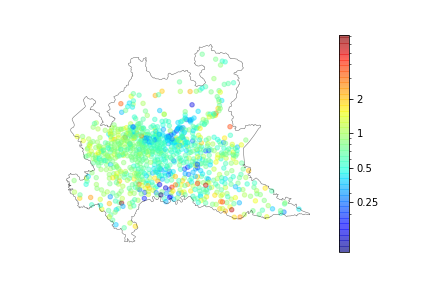} 
    \end{subfigure}
    \end{center} 

\raggedright {\footnotesize \emph{Note:} distribution of estimated per-municipality
$R_0$, computed as $\frac{\beta}{\gamma}$.}
\end{figure}

Figure \ref{fig:r0-dist} shows the distribution of the estimated values of $R_0$: the mean
estimated value is 0.83 
(0.85 when weighted on population), 
while the median is 0.70
A strong heterogeneity (which can
be partly attributed to statistical noise -- several municipalities count only a
few cases
each) can be observed across
municipalities:
in what follows, unless differently specified, we trim data by
dropping 0.5\% of outliers on each side of the distribution of $R_0$, hence
reducing to 903 municipalities. 
In only few of these (179) 
the value of $R_0$ appears to be larger
than the critical threshold of 1: that is, in the vast majority of
municipalities, the outbreak is expected to spontaneously extinguish without
requiring herd immunity.

\begin{table}
\begin{center}
\caption{Main results}
\label{tab:main}
\begin{tabular}{@{\extracolsep{5pt}}lcccc}
\\[-1.8ex]\hline
\\[-1.8ex] & (1) & (2) & (3) & (4) \\
\hline \\[-1.8ex]
 Intercept & 0.920$^{***}$ & 0.921$^{***}$ & 0.906$^{***}$ & 0.823$^{***}$ \\
  & (0.024) & (0.025) & (0.025) & (0.016) \\
 cases\textperthousand & -0.023$^{***}$ & -0.023$^{***}$ & -0.032$^{***}$ & \\
  & (0.004) & (0.004) & (0.005) & \\
 cases & & & -0.000$^{}$ & -0.001$^{***}$ \\
  & & & (0.000) & (0.000) \\
 population & & -0.000$^{}$ & & 0.000$^{***}$ \\
  & & (0.000) & & (0.000) \\
 population$^{-1}$ & & & 180.274$^{***}$ & \\
  & & & (57.829) & \\
\hline \\[-1.8ex]
 Observations & 903 & 903 & 903 & 903 \\
 R${2}$ & 0.036 & 0.036 & 0.048 & 0.015 \\
\hline
\hline \\[-1.8ex]
\end{tabular}\\
\raggedright{\footnotesize \emph{Note:} dependent variable $R_0$.
$^{*}$p$<$0.1; $^{**}$p$<$0.05; $^{***}$p$<$0.01}
\end{center}
\end{table}

Table \ref{tab:main} presents the results of the regression analysis.
We see a negative and strongly significant relationship between the initial
number of cases per one thousand inhabitants and the estimated $R_0$ (column (1));
this relationship is
robust to controlling for population size (column (2)), and to both the absolute
number of cases and the inverse of population size (column (3)), i.e., a full interaction
model where the the per capita count represents the interaction term
\citep{Kronmal}.
The coefficient for the per capita number of cases can be
interpreted as the
reduction in $R_0$ resulting from an increase of one case per one
thousand individuals in the prevalence of the outbreak.
The value of -0.023 observed in column (2), which we consider as our baseline
specification, 
indicates a sizeable effect:
for reference, given that the prevalence in Milan as of March 25 was of
around $1.7$\textperthousand, the above mentioned result suggests that had it been
$2.7$\textperthousand, the average $R_0$ would have been around
0.88 
instead than the observed 0.90. 
The same negative and strongly significant effect is observed if we focus on the
absolute number of cases as explanatory variable, controlling for the population
size (column (4)).

It should be noted that any intrinsic characteristic of municipalities -- such
as demography, location, structure of the economy -- which might explain a
larger outbreak size should also favor a larger $R_0$ \citep{demography}.
Thus, controlling for such characteristics is expected to further reduce the
coefficient for \emph{cases\textperthousand}.

\subsection{Interpretation}
\label{sec:interp}

There are a few reasons that might explain why a larger
outbreak should result in a \emph{subsequent} lower $R_0$.

The first might be related to herd immunity, by which areas where the outbreak
is initially more present have less scope for further spread because a large
share of individuals have already caught, and possibly developed immunity to,
the virus.
This is in principle not a problem of our approach, as the SIR model accounts
for this effect and should estimate an $R_0$ net of it -- in other terms, $R_0$
describes the evolution of the outbreak in an hypothetical situation in which
the pool of susceptible individuals is never reduced.
However, the problem might still arise if the count data employed severely
underestimate the actual spread of the virus:
the number of positive cases could actually be
much larger than the detected one, leading to an estimated $R_0$ lower than the
real one because of the undetected effect of herd immunity in reducing the
rate of contagion.

The underestimation of infected population might also suggest an alternative
explanation of
the result related to \emph{test capacity}:
to the extent that a lower detected prevalence reflects a lower ability of
authorities to identify infected individuals, it should then correlate with a
lower ability to isolate, hospitalize and cure them, and hence to a
faster outbreak growth.

A third, \emph{social}, explanation is instead that wherever the local
population is aware of a
larger prevalence of the disease, it reacts by changing its behavior towards a
stricter application of social distancing rules, thus leading to a lower $R_0$.
In what follows, we provide evidence in favor of this hypothesis.

We start by analyzing the first possible explanation: several sources have
argued that the actual size of the infected population might lie between four
and ten times the official reported numbers.
In the most affected muncipalities in our sample during the period analyzed,
57 infections per 
one thousand inhabitants have been recorded, and according to the most
pessimistic estimates this
would mean that up to $57\%$ 
of the population was infected.
While most municipalities have a number of recorded cases per
one thousand inhabitants which is orders of magnitude lower,
to avoid the possibility that an even partial herd immunity effect
might be driving the results, we re-estimate our main model on subsamples of
municipalities according to their initial number of cases per capita.
Specifically, we split the sample according to quartiles of cases per capita
on March 25.
The results, presented in columns (1) to (4) of Table \ref{tab:sec}, show that our 
findings are
not driven by herd immunity, as the coefficient for \emph{cases\textperthousand}
is negative in each quartile.
The absolute value of such coefficient is actually much larger
for municipalities with a low prevalence than for those with a higher
prevalence, and is strongly significant in the first two quartiles, hence
including municipalities with 2 cases 
per one thousand inhabitants or less.

We then consider the second possible explanation: that a lower \emph{detected}
prevalence signals a lower detection ability, and that this naturally correlates
with lower ability to track and quarantine infected subjects, hence raising the
subsequent rate of diffusion.
In order to disentangle this \emph{test capacity} explanation from the third,
\emph{social}, one, we sketch two simple models of how these would be expected
to affect $R_0$.

Let us represent with $u_{t}$ the unknown real number of infected subjects per
one thousand inhabitants, at time $t$ in a given municipality, and with $i_t$
the corresponding known number.
We are interested in the extent to which unidentified infected subjects (which
are $u_t - i_t$ cases per one thousand inhabitants) will raise the $R_0$ for the
municipality in the subsequent period.
More specifically, we can assume that identified and unidentified patients form
two different pools of infected subjects and that the latter has a much
higher $\beta$ -- probability of infecting susceptible individuals -- that
leads to a corresponding higher $R_0$.
Since $\beta$ enters linearly in $R_0$ -- and assuming for simplicity that
$\gamma$ is constant -- the relationship between $u_t - i_t$ and $R_0$ would be
expected to be \emph{linear}.
Moreover, it is well known that not only identified patients are subject to a
stronger form of isolation, but also close contacts of such patient (some of which
are not infected) are
recommended to self-quarantine: this does not happen in municipalities with a
larger number of undetected cases, which implies that the effect of each
unidentified patient should be \emph{more} than linear in increasing the $R_0$.
This would imply a linear or concave relationship between  
\emph{cases\textperthousand} and $R_0$.

Vice-versa, any \emph{social} explanation is based on the assumption that
inhabitants react to the
news of the cases in their municipality.
Given any concave function describing this reaction, a same increase in
per capita cases will be perceived as more important if the initial number of
cases is lower.
That is, we can expect inhabitants of two towns with respectively 1 and 11 known
cases per one thousand inhabitants to differ in their compliance with social
distancing prescriptions more than inhabitants of two towns with respectively
20 and 30 known cases per one thousands inhabitants: a same difference of
one percentage point in prevalence will have a weaker effect on people behavior 
were prevalence is higher.
This alternative explanation predicts a \emph{convex}
relationship (given the negative sign) between \emph{$cases$\textperthousand}
and $R_0$.

\begin{table}
\begin{center}
\caption{Additional specifications}
\label{tab:sec}
\begin{tabular}{@{\extracolsep{5pt}}lccccc}
\\[-1.8ex]\hline
\\[-1.8ex] & \multicolumn{1}{c}{Q 1} & \multicolumn{1}{c}{Q 2} & \multicolumn{1}{c}{Q 3} & \multicolumn{1}{c}{Q 4} & \multicolumn{1}{c}{Full}  \\
\\[-1.8ex] & (1) & (2) & (3) & (4) & (5) \\
\hline \\[-1.8ex]
 Intercept & 1.071$^{***}$ & 1.290$^{***}$ & 0.899$^{***}$ & 0.760$^{***}$ & 0.992$^{***}$ \\
  & (0.066) & (0.185) & (0.206) & (0.098) & (0.033) \\
 cases\textperthousand & -0.098$^{**}$ & -0.125$^{**}$ & -0.033$^{}$ & -0.004$^{}$ & -0.050$^{***}$ \\
  & (0.045) & (0.057) & (0.039) & (0.009) & (0.009) \\
 cases\textperthousand$^2$ & & & & & 0.002$^{***}$ \\
  & & & & & (0.001) \\
 population & 0.000$^{}$ & -0.000$^{}$ & 0.000$^{}$ & -0.000$^{}$ & -0.000$^{}$ \\
  & (0.000) & (0.000) & (0.000) & (0.000) & (0.000) \\
\hline \\[-1.8ex]
 Observations & 226 & 226 & 225 & 226 & 903 \\
 R${2}$ & 0.021 & 0.027 & 0.005 & 0.008 & 0.047 \\
\hline
\hline \\[-1.8ex]
\end{tabular}\\
\raggedright{\footnotesize \emph{Note:} dependent variable: estimated $R_0$.
Columns (1) to (4): model restricted to municipalities with a number of cases
per thousand inhabitants in the interval
(0.278, 2.177],
(2.177, 4.124],
(4.124, 6.277] and
(6.277, 30.973], respectively. 
Column (5): full sample.
$^{*}$p$<$0.1; $^{**}$p$<$0.05; $^{***}$p$<$0.01}
\end{center}
\end{table}

To disentangle between the \emph{test capacity} and the \emph{social} 
explanation, we enrich our
basic model by introducing a quadratic term in \emph{cases\textperthousand}. 
This is done in
column (5) of Table \ref{tab:sec}. We see that the quadratic term has a
positive sign and is strongly significant, while the sign of the linear term is
still negative and has increased in absolute terms.
Hence, while this does not allow us to exlude that the other explanations
might play a role, we can conclude that the \emph{social} explanation is the
main driver of the negative relationship between \emph{cases\textperthousand} 
and $R_0$.

\citep{jones2020optimal} describes two possible opposite reactions to the
COVID-19 oubreak: a \emph{precautionary} attitude that leads to a stricter
adherence to guidelines, and a ``fatalism effect'' according to which an
individual who is
more likely to be infected in the future ``\emph{reduces her incentives to be
careful today}''.
Our results provide strong evidence in favor of the first mechanism.

\subsection{Sensitivity tests}
\label{sec:sensitivity}

\begin{figure}
    \begin{center}
    \caption{}
    \begin{subfigure}[b]{8cm}
        \includegraphics[width=7cm]{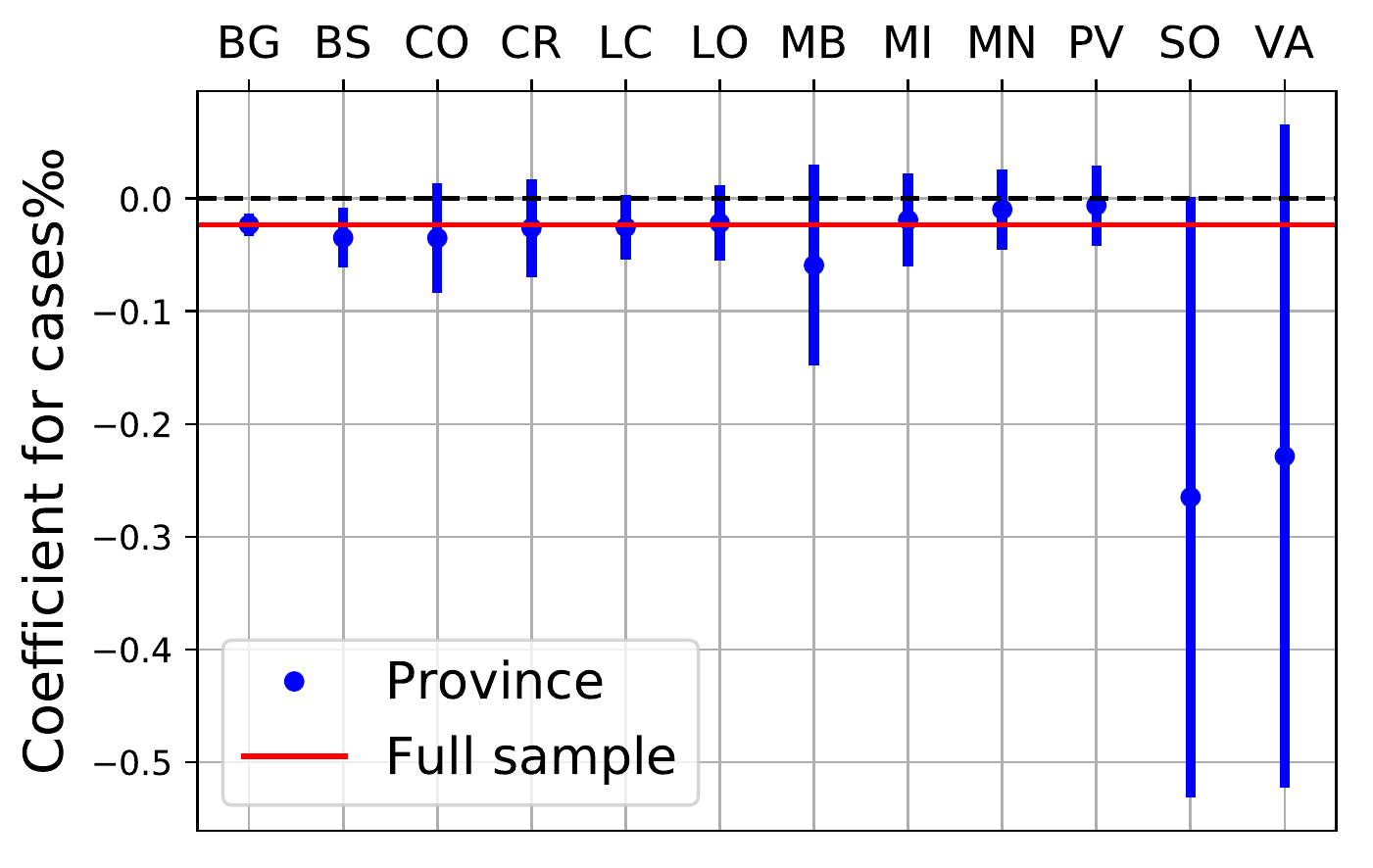}
    \caption{Disaggregated estimation across provinces}
    \label{fig:rob-prov}
    \end{subfigure}
    ~
    \begin{subfigure}[b]{8cm}
        \includegraphics[width=7cm]{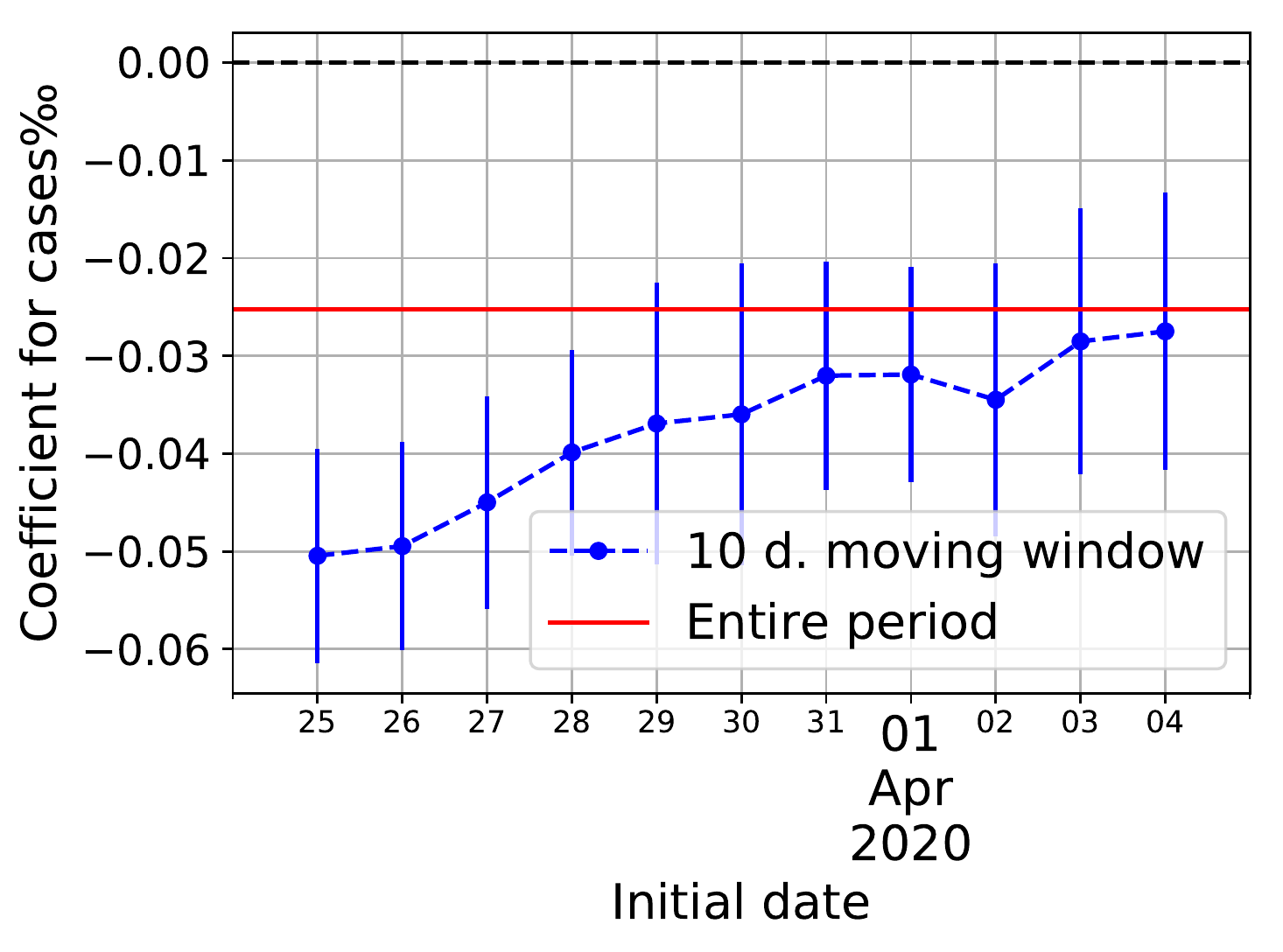}
    \caption{Estimation over 10-days moving windows}
    \label{fig:rob-window}
    \end{subfigure}
    \end{center}
\raggedright {\footnotesize
\emph{Note:} estimates are run controlling for population, as in
column (2) of Table \ref{tab:main}; the red line denotes the corresponding
coefficient estimated on the entire sample under analysis.}
\end{figure}

In addition to the quantile analysis previously described, we verify that our
main result also holds consistently across the 12 provinces (lower level
administrative regions) of which Lombardy is composed.
Results are displayed in Figure \ref{fig:rob-prov}. We see that, for each 
province, the effect of \emph{cases\textperthousand} on $R_0$ is negative:
although the small sample size results in only few provinces reaching
statistical significance, it is clear that no specific area of Lombardy is
alone responsible for our findings.

In order to verify that our results do not strictly depend upon the period
considered,
we replicate our analysis over different 10-days moving windows within our
period of analysis.
For each subperiod, we fit the $R_0$ for each municipality and regress it
on the number of cases per thousand individuals at the beginning of that
subperiod.
In accordance with the selection procedure described in Section \ref{sec:methods}, 
we reduce this analysis to the 713 municipalities that feature at least one case
on March 25
and, \emph{in each window}, have new cases recorded in at least two dates.
The results are shown in Figure \ref{fig:rob-window}. For comparability, we
also display the value of the coefficient estimated for the entire time period
on the same restricted sample of 713 municipalities.
We find that the effect of interest is robust, that is, the coefficient for the
\emph{cases\textperthousand} is consistently negative and strongly significant
for each subperiod.
Its absolute value is significantly decreasing over time; that is, the
effect of the number of cases on the $R_0$ in the following days appears to be
stronger in the earlier days of the outbreak.
While there might be multiple explanations for this, we only remark that the
rate of growth of the epidemic has been consistently decreasing:
whether individual behavior reacts not just
to outbreak size, but also to its \emph{change} over time, is an issue for
further research.

Finally, we verify that all results reported in Table \ref{tab:main}, including
statistical significance, are virtually unchanged if we do not trim the data as
previously described.

\subsection{Prediction of outbreak duration}

While an accurate predition of the date of extinction of the outbreak
deserves more sophisticated epidemiologic models
\citep{guzzetta2020potential,FBK}
that are out of the scope of
the present paper, we can analyze to some extent the relationship between the
predicted
date of extinction and the number of initial cases.

In general, the relationship between infected population at time $t$ and
expected date of extinction of the outbreak within a SIR model depends on the
size of $R_0$:
if the latter is
smaller than 1 -- i.e., the outbreak is spontaneously slowing -- then a smaller
outbreak will extinguish sooner; vice-versa, if $R_0>1$, a larger outbreak will
sooner reach a level of herd immunity, and hence die out.

Since, according to our data, most municipalities in Lombardy 
display an $R_0 < 1$, we focus on this case.
While for a same level of $R_0$ we expect the predicted time to extinction to
increase with the initial outbreak size, the fact that the $R_0$ is
\emph{negatively} related to initial outbreak size -- and that a lower $R_0$
leads to a quicker extinction -- leaves theoretically undetermined the
relationship between initial outbreak size and duration of the outbreak.

\begin{table}
\begin{center}
\caption{Outbreak size and expected outbreak duration}
\label{tab:days}
\begin{tabular}{@{\extracolsep{5pt}}lcccc}
\\[-1.8ex]\hline
\\[-1.8ex] & \multicolumn{1}{c}{Full} & \multicolumn{1}{c}{$R_0$<1} & \multicolumn{1}{c}{Full} & \multicolumn{1}{c}{$R_0$<1}  \\
\\[-1.8ex] & (1) & (2) & (3) & (4) \\
\hline \\[-1.8ex]
 Intercept & 214.957$^{***}$ & 56.873$^{***}$ & 317.477$^{***}$ & 142.455$^{***}$ \\
  & (24.500) & (13.010) & (17.771) & (11.028) \\
 cases\textperthousand & 7.032$^{*}$ & 8.198$^{***}$ & & \\
  & (3.999) & (2.082) & & \\
 cases & & & -0.339$^{}$ & 0.114$^{}$ \\
  & & & (0.353) & (0.201) \\
 population & -0.000$^{}$ & 0.000$^{}$ & 0.001$^{}$ & 0.000$^{}$ \\
  & (0.000) & (0.000) & (0.001) & (0.000) \\
\hline \\[-1.8ex]
 Observations & 903 & 729 & 903 & 729 \\
 R${2}$ & 0.004 & 0.021 & 0.002 & 0.005 \\
\hline
\hline \\[-1.8ex]
\end{tabular}\\
\raggedright{\footnotesize \emph{Note:} dependent variable: days to expected
outbreak extinction.
``Extinction'' is defined as reaching 0.1 cases per one thousand inhabitant in
columns (1) and (2), 0.1 cases in columns (3) and (4).
$^{*}$p$<$0.1; $^{**}$p$<$0.05; $^{***}$p$<$0.01}
\end{center}
\end{table}

In order to shed light on this indeterminacy, we proceed to simulating the SIR
model for each municipality until the predicted size of the infected population
decreases below either (i) 0.1 cases for one thousands inhabitants
or (ii) 0.1 cases\footnote{SIR models by design tend to
0 infected subjects only for $t \to \infty$ and different authors pick different
thresholds as denoting outbreak extinction. Notice that the most appropriate
value crucially depends also on the extent to which the outbreak is
underestimated by available data.}
and we consider the number of periods elapsed as the \emph{outbreak duration}.
We then regress the outbreak duration, defined in these two different ways, over
the initial (i) number of cases per
capita  (columns (1) 
and (2) of Table \ref{tab:days}) and (ii) absolute number of cases
(columns (3) and (4) of Table \ref{tab:days}), respectively.

Results from Table \ref{tab:days} show that the relationship between outbreak
size and extinction date is non-trivial.
First, the relatively few municipalities with $R_0>1$ do influence significantly
the
results -- as already discussed, the expected effect of outbreak size for a same
$R_0$ is reversed in such cases.
Second, if we restrict to $R_0<1$,
the relationship is positive and significant when reasoning in per capita
terms, but not in absolute terms.
It should also be mentioned that the results depend on the thresholds adopted
in the definition of outbreak extinction.
In general, given that $R_0<1$ determines the exponential decay, a lower
threshold will mean that the date of extinction is further
away for municipalities with a relatively low number of cases and relatively
high $R_0$.

Summing up, the results of predicting the extinction date are to be interpreted
as cautionary: municipalities with smaller outbreaks might get rid of them
sooner than others with more infected individuals (column (2) of Table
\ref{tab:days}), but this result does not generalize to the absolute outbreak
size (columns (3) and (4) -- the former even featuring a negative sign).
Plans for a gradual exit from lockdown should take into account that the
relationship between outbreak size and expected outbreak duration is difficult
to pinpoint -- as well as the possibility that a larger outbreak
might bring the population closer to herd immunity, making it more resistent to
a new outbreak.

\section{Conclusion}

We show that in Lombardy, during a lockdown, the basic reproduction number for
COVID-19 reacts negatively
to the initial size of an outbreak at the municipality level, an
effect which cannot be explained by the population having reached herd immunity.
Limited test capacity -- and hence a limited 
ability by health authorities to isolate and treat affected individuals --
appear to have
at most a marginal role in explaining our result.
Instead, we show that the population's behavior is key to slowing down the contagion
and in particular that information about local outbreaks impacts on diffusion
rates.
This effect is consistent across all provinces and it is robust to the
sample period considered.

The fact that the effect is particularly strong in municipalities characterized 
by a smaller outbreak suggests that individuals react more
strongly to the first few cases.
This aspect is confirmed by the convex relationship we find between the initial
size of
the outbreak and the $R_0$: the marginal effect on 
behavior of each new case
seems to decrease in the number of cases.

Our results provide evidence in favor of a \emph{precautionary} rather than
\emph{fatalistic} individual attitude towards the outbreak.
They call for considering the population as an
integral part of the decision making process, and for a timely and transparent
provision of epidemiologic data.

\newpage

\addcontentsline{toc}{section}{References}
\bibliographystyle{chicago}
\bibliography{bg_covid_lombardy}

\newpage

\section*{Supplementary information}
\subsection*{Optimization procedure}

For simplicity, the procedure for fitting the SIR model is implemented over the
parameters $R_0$ and
$\gamma$ rather than $\beta$ and $\gamma$, where $R_0 = \frac{\beta}{\gamma}$.
For each parameter (including the initial values $I = \hat I$ and $R = \hat R$),
the procedure
is initialized by deriving reasonable values based on tuning the model to
regional aggregated data.

Then the procedure works as follows ($i$ denotes an iteration, $\delta_{\pi,1}$
is set to $0.2$ for each parameter $\pi$):
\begin{enumerate}
\item given the current value of a parameter $\pi_i$, compute values of the
parameter $\pi_{i-l} = \pi_i \times (1-\delta_{\pi,i})$ and
$\pi_{i,r} = \pi_i \times (1+\delta_{\pi,i})$ (the \emph{right} and \emph{left
candidate values} for parameter $\pi$),
\item compare count data and the simulation obtained with each of the three
candidates $\pi_{i,l}$, $\pi_{i}$ and $\pi_{i,r}$ by computing the sum of
squared residuals,
\item select the candidate value which results in the smallest error as new
parameter value $\pi_{i+1}$,
\item if the value did \emph{not} change (that is, $\pi_{i+1} = \pi_i$),
set $\delta_{\pi,i+1} = \delta_{\pi,i} \cdot 0.75$; otherwise, leave
$\delta_{\pi,i+1} = \delta_{\pi,i}$,
\item repeat steps from 1 to 4 for each parameter $\pi \in \{R_0, \gamma, \hat
I, \hat R\}$,
\item repeat steps from 1 to 5 until $\delta_{\pi,i} < 0.001$ for each
parameter.
\end{enumerate}

\end{document}